# Polarity-dependent dielectric torque in nematic liquid crystals

Mingxia Gu, Sergij V. Shiyanovskii and Oleg D. Lavrentovich

Chemical Physics Interdisciplinary Program, Liquid Crystal Institute, Kent State

University, Kent, Ohio 44242

**ABSTRACT**

The dielectric dispersion in the uniaxial nematic liquid crystals affects the switching dynamics of the director, as the dielectric torque is determined by not only the present values of the electric field and director but also by their past values. We demonstrate that this "dielectric memory" leads to an unusual contribution to the dielectric torque that is linear in the present field and thus polarity-sensitive. This torque can be used to accelerate the "switch-off" phase of director dynamics.

The reorientation of nematic liquid crystal (NLC) molecules by an electric field $\mathbf{E}$ is used in many electrooptical applications, most notably the Liquid Crystal displays (LCDs) [1]. Tremendous efforts have been put into improving the performance of LCDs, especially their switching time. There are two phases in electric switching of a LCD. The fast "active" phase of "switch on" is driven by an applied voltage $U$ with the characteristic time $\tau_{on} \approx \gamma_1 d^2 / (\varepsilon_0 |\Delta\varepsilon| U^2)$ (here $\varepsilon_0$ is the electric constant, $d$ is the cell thickness, $\gamma_1$ is the rotational viscosity, $\Delta\varepsilon = \varepsilon_\parallel - \varepsilon_\perp$ is the dielectric anisotropy, $\varepsilon_\parallel$ and $\varepsilon_\perp$ are the principal dielectric permittivities referred to the director $\hat{\mathbf{n}}$) [1]; $\tau_{on}$ can be decreased by increasing $U$. Director reorientation in the "switch off" phase is a "passive" process driven by relaxation of elastic distortions with switch-off time $\tau_{off} \approx \gamma_1 d^2 / (\pi^2 K)$ [1] that depends on the NLC properties such as $\gamma_1$ and the elastic constant $K$, but not on the (pre)applied electric field. This consideration is based on a classic picture of a NLC as a medium with no dielectric dispersion and instant dielectric response; the dielectric torque $\mathbf{M}_d = \varepsilon_0 \Delta\varepsilon (\mathbf{E} \cdot \hat{\mathbf{n}}) \mathbf{E} \times \hat{\mathbf{n}}$ is quadratic in $\mathbf{E}$ and is determined by the present values of $\mathbf{E}$ and $\hat{\mathbf{n}}$ [1]. Frequency dependence of dielectric permittivity changes this picture, as $\mathbf{M}_d$ becomes dependent not only on the present $\mathbf{E}$ and $\hat{\mathbf{n}}$, but also on their past values [2, 3]. This "dielectric memory effect" (DME) has been studied for so-called dual frequency NLCs in which $\Delta\varepsilon$ changes sign with the frequency $f$ [2-4]. In this work, we demonstrate a very unusual consequence of DME, namely, an existence of a "memory" dielectric torque that is *linear* (rather than quadratic) in the present $\mathbf{E}$. The direction of the torque can be controlled by the polarity of $\mathbf{E}$, regardless of the sign of $\Delta\varepsilon$. This feature allows one to design a situation when the "switch-off" phase can be accelerated by a properly chosen back edge of the electric pulse and is no longer a "passive" process.

**Theory.** Due to dielectric relaxation, electric displacement $\mathbf{D}(t)$ and the torque $\mathbf{M}_d(t) = \mathbf{D}(t) \times \mathbf{E}(t)$ depend on both the present $\mathbf{E}(t)$, and the past field $\mathbf{E}(t')$, $-\infty < t' \leq t$ [2,3]. Many NLCs experience only a single relaxation process at $f < 10$ MHz that can be described by the Debye model:

$$\varepsilon_\parallel(f) = \varepsilon_{h\parallel} + \frac{\varepsilon_{l\parallel} - \varepsilon_{h\parallel}}{1 - i2\pi f \tau}, \quad \varepsilon_\perp(f) = \varepsilon_\perp = const, \tag{1}$$

where "*l*" and "*h*" refer to the low and high $f$, $\tau$ is the dielectric relaxation time. We focus on two Debye type materials with different (but $f$-independent) signs of $\Delta\varepsilon$ with dielectric relaxation in the kHz region convenient for experimental studies (although the consideration is applicable for other parts of the spectrum). The negative $\Delta\varepsilon < 0$ NLC was obtained by mixing 60.6 wt % MLC-7026-100 (EM Industries) and 39.4 wt % 2F-3333 (Rolic Technologies); the positive $\Delta\varepsilon > 0$ NLC was a mixture of 20.0 wt % of pentylcyanobiphenyl (5CB, EM Industries) and 80.0 wt % 2F-3333, Fig. 1. The dielectric permittivities were measured using Schlumberger 1260 impedance/gain-phase analyzer.

The director dynamics is determined by the balance of the dielectric $\mathbf{M}_d$, viscous $\mathbf{M}_v$, and elastic $\mathbf{M}_e$ torques through the Ericksen-Leslie equation $\mathbf{M}_d + \mathbf{M}_v + \mathbf{M}_e = 0$. For a flat cell with plates along the *x-y* plane, $\mathbf{E}(t) = E_z(t)\hat{\mathbf{z}}$, $\hat{\mathbf{n}}(t)$ in the *x-z* plane depending only on *z*, the only non-zero components of torques are along the y-axis. The dielectric torque is [2]:

$$M_d(t) = \varepsilon_0 E(t) \sin\theta(t) \left[ \Delta\varepsilon_h E(t) \cos\theta(t) + \frac{\varepsilon_{l\parallel} - \varepsilon_{h\parallel}}{\tau} \int_{-\infty}^{t} \exp\left(-\frac{t-t'}{\tau}\right) E(t') \cos\theta(t') dt' \right], \tag{2}$$

where $\theta(t)$ is the angle between $\hat{\mathbf{n}}(t)$ and $\mathbf{E}(t)$. For $\varepsilon_{l\parallel} = \varepsilon_{h\parallel} = \varepsilon_{\parallel}$, Eq. (2) recovers the classic "instantaneous" dielectric response theory.

Consider the torque balance in response to a voltage changes over a short time interval between $t = 0$ and $t \sim \tau$. This interval is short enough to assume the changes of $\theta(t)$ small. This allows us to approximate as time independent the following three quantities: (a) $M_d$ with $\theta(z,t) \approx \theta(z, t = 0) = \theta_0(z)$ (provided $\theta_0(z) \neq 0, \pi/2$), (b) the elastic torque $M_e(z,t) = M_e(z)$ and (c) the spatial non-uniformity of the electric field, $E_z(z,t) = g(z)U(t)$, where $g(z)$ is a proportionality factor. We neglect the back-flow effect, thus $M_v(t) = \gamma_1 d\theta(t)/dt$. Under these assumptions, the solution of Ericksen-Leslie equation $\gamma_1 d\theta(t)/dt = -M_d - M_e$ that describes the time evolution of the system, is $\theta(t) = \theta_0(z) - \Theta(z)Q + M_e(z)t$, where $\Theta(z) = \varepsilon_0(\varepsilon_{l\parallel} - \varepsilon_{h\parallel})\tau U_0^2 g^2(z)\sin 2\theta_0(z)/\gamma_1$ and $Q$ is the normalized integrated dielectric torque:

$$Q = \tau^{-1}\int_0^t \left\{ \xi\, u^2(t') + u(t')\left\{ u_{mem}\exp(-t'/\tau) + \tau^{-1}\int_0^{t'} \exp\left[-(t'-t'')/\tau\right]u(t'')dt'' \right\} \right\} dt'. \quad (3)$$

Here $\xi = (\varepsilon_{h\parallel} - \varepsilon_{\perp})/(\varepsilon_{l\parallel} - \varepsilon_{h\parallel})$, $u(t) = U(t)/U_0$ is the normalized voltage with $U_0 = U(t \to 0^-)$ being the voltage that acts just before the "switch-off" moment $t = 0$, $u_{mem} = \tau^{-1}\int_{-\infty}^0 \exp(t'/\tau)u(t')dt'$ is the memory term caused by the "past" field.

To optimize $Q$, we apply a direct variational method. Integrating Eq. (3) with a decaying exponential probe function $u(t) = a\,\mathrm{Exp}(-\Gamma t/\tau)$, where $a$ and $\Gamma > 0$ are two variational parameters, one obtains:

$$Q = \frac{a\left[2u_{mem}\Gamma + a(1+\xi+\xi\Gamma)\right]}{2\Gamma(1+\Gamma)} - \frac{a^2(1+\xi-\xi\Gamma)}{2\Gamma(\Gamma-1)}e^{-2\Gamma t/\tau} + \frac{a\left[u_{mem}(\Gamma-1)+a\right]}{\Gamma^2-1}e^{-(1+\Gamma)t/\tau}. \quad (4)$$

The last expression will be used to fit the data below. It is instructive to qualitatively discuss the basic feature of Eq. (4). The first term dominates for large $t > 5\tau$ and determines the saturated value of $Q$. Its extremum $Q_e = u_{mem}^2\left\{\left[(1+\xi)\xi\right]^{1/2} - \xi - 1/2\right\}$ is reached for $\Gamma_e = (1+\xi^{-1})^{1/2}$ and $a_e = u_{mem}(1-\Gamma_e)$. It is easy to see that $Q_e < 0$ for the positive NLCs in which $\xi > 0$, while $Q_e > 0$ for the negative NLCs in which $\xi < -1$. The $Q_e$ is opposite in sign to the dielectric torque in the switch on phase, thus it can accelerate the director relaxation in the switch off phase.

The theory can be qualitatively explained as follows. Consider first a NLC with $\Delta\varepsilon > 0$ in a planar cell. A positive dc field $E_z > 0$ reorients $\hat{\mathbf{n}}$ towards the z-axis. $E_z$ also induces a dipole moment density $\mathbf{p}$ with the components $p_\perp = \varepsilon_\perp E_z \sin\theta$ and $p_\parallel = \varepsilon_{h\parallel}E_z\cos\theta + p_{mem}$, perpendicular and parallel to $\hat{\mathbf{n}}$, respectively. Here $p_{mem} \propto u_{mem}$ is the "memory" contribution that saturates to the value $p_{mem} = (\varepsilon_{l\parallel} - \varepsilon_{h\parallel})E_z\cos\theta$ after the dc field $E_z$ acted for a sufficiently long time $> \tau$ ($u_{mem} \to 1$). Note that $p_{mem} > 0$ and $E_z > 0$ are of the same sign. When the field is switched off at $t = 0$, $p_{mem} > 0$ does not disappear instantaneously, but decays with a characteristic time $\tau$. If within the interval $0 < t \leq \tau$, one applies a new electric pulse of *the opposite polarity*, $E_z < 0$, then this field would interact with the decaying $p_{mem} > 0$ to assist the reorientation towards the planar state, $\theta \to \pi/2$.

In the homeotropic cell with a negative NLC, the field $E_z > 0$ at $t < 0$ also induces $p_{mem} > 0$ (of the same polarity). If within the interval $0 \leq t \leq \tau$ one applies a new voltage pulse of *the same* polarity, $E_z > 0$, then this field will couple to $p_{mem} > 0$ to assist the director reorientation into the homeotropic state, $\theta \to 0$.

If the NLC were not dispersive, any field-induced polarization would relax instantaneously, $p_{mem} = 0$, $\varepsilon_{l\|} = \varepsilon_{h\|}$ and $\tau = 0$, and the effect would not be observed. Therefore, the classic theory with an instantaneous dielectric response cannot predict the phenomenon we describe. Below we present experimental verifications of the effect.

**Experiment.** We used homeotropic and planar (with a small pretilt of ~ $1^o$) cells (EHC Ltd.) comprised of glass substrates with indium tin oxide electrodes of area $10 \times 10 \text{ mm}^2$; $d = 14.4 \ \mu\text{m}$ for the homeotropic cell and $d = 20.6 \ \mu\text{m}$ for the planar cell. The field-induced director dynamics was monitored by measuring the He-Ne laser ($\lambda = 633 \text{ nm}$) light transmission $I(t) = A \sin^2(\Phi/2)$ through the cells placed between two crossed polarizers [1]. The phase retardation $\Phi$ depends on $\theta$; for small variations of $\theta$, the retardation change is linear in $Q$, $\Phi - \Phi_0 \approx \rho Q$, where

$$\Phi_0 = \frac{2\pi n_o}{\lambda} \int_0^d \left\{ \frac{n_e}{\tilde{n}(z)} - 1 \right\} dz, \ \rho = \frac{\pi n_o n_e \left(n_e^2 - n_o^2\right)}{\lambda} \int_0^d \frac{\Theta(z) \sin 2\theta_0(z)}{\tilde{n}(z)^3} dz, \qquad (5)$$

and $\tilde{n}(z) = \left[ n_e^2 \sin^2 \theta_0(z) + n_o^2 \cos^2 \theta_0(z) \right]^{1/2}$. The coefficient $A$ is close to the intensity $I_0$ of the impinging light for the planar cell when the rubbing direction is at $45^o$ with respect to the polarizers [1]. In the homeotropic cell, $A \approx I_0/2$ because the applied field creates random azimuthal orientation of $\hat{\mathbf{n}}$ with numerous umbilics, 10-100 within the area probed by the beam. These defects randomize the director field in the plane of the cell thus assuring reproducibility of the experiment. The umbilics relax much slower (seconds and minutes [5]) than the duration of our experiments ($< 0.5$ ms). We used TIA-500S-TS photodetector (Terahertz Technologies) and Tektronix TDS 210 oscilloscope to measure $I(t)$.

The driving pulses were produced by WFG500 wave-form generator (FLC Electronics); the maximum rate was 240 V/μs. To test the switch-off dynamics, we used two different profiles for the pulse's back edge: (i) an instantaneous back edge (in practice ~1 μs in duration because of the finite voltage change rate); (e) an exponentially decaying back edge $u(t) = a \, \text{Exp}(-\Gamma t/\tau)$.

To drive the homeotropic cell with a negative NLC, we first apply a square 100 V dc pulse of duration 225 μs, much longer than $\tau = 33$ μs, Fig.1(a), so that there is enough time to produce the saturated "memory" dipole moment ($u_{mem} \to 1$). This pulse is switched off by an instantaneous back edge (i) or by three different exponential edges with $\Gamma = 0.45$ and: (e1) positive polarity, $a = 0.5$; (e2) $a = 0.87$; (e3) negative polarity, $a = -0.5$; Fig. 2. The optical response is different in all four cases. In the case (i), $\hat{\mathbf{n}}$ reorients slowly toward the homeotropic state $\theta \to 0$, as evidenced by the decrease in $I(t)$ in Fig. 2 inset. The pulse (e1) produces much faster reorientation (r-e1), despite the fact that $U$ decreases less abruptly as in case (i). The shape of the pulse (e1) is close to the optimum, as any departure from the pre-selected $a = 0.5$ and $\Gamma = 0.45$ causes a slower or even a non-monotonous response, as in (e2) case. The linear $\mathbf{E}$-dependence of the "memory" torque is well illustrated by the response to pulses (e1) and (e3) that are identical in amplitude and duration and differ only in polarity: (e1) drives $\hat{\mathbf{n}}$ toward $\theta = 0$ while (e3) continues to drive $\hat{\mathbf{n}}$ toward $\theta = \pi/2$. After a sufficiently long time, the NLC relaxes to the same homeotropic state with $I = 0$ for all pulses, Fig. 2.

The different scenarios can be fitted by the model above. To fit the data, we used the approximation $\Phi - \Phi_0 \approx \rho Q(t)$, Eq. (5); the only fitting parameter is $\rho$, as $Q(t)$ is determined by the experimental values of $a$ and $\Gamma$, Eq. (4). We first fit the response curve (r-e1) using $a = 0.5$, $\tau = 33\ \mu s$ and $\Gamma = 0.45$ and find $\rho = 2.33$. With this value of $\rho$, and with the experimental $\tau$, $a$ and $\Gamma$, the model (4) reproduces the measured curves (r-e2) and (r-e3) very well with no fitting parameters, Fig. 2.

For the planar cell with the positive NLC, we used a 100 V dc pulse of duration $120\ \mu s$, Fig. 3. The back edge was either (i) instantaneous or (e) exponential $u(t) = a\mathrm{Exp}(-\Gamma t/\tau)$, with a negative $a = -1$, $\tau = 20\ \mu s$ and two different decaying speeds, $\Gamma = 2$ (e1) and $\Gamma = 0.75$ (e2). The optical responses, Fig. 3(a), share one common feature, namely, a universal decay that does not depend on the details of the back edge. This universal decay has a characteristic time of 0.5 ms that is much shorter than the elastic relaxation time, $\tau_{off} \approx 0.4\ \mathrm{s}$ for a typical $\gamma_1 \sim 0.1\ \mathrm{kg\ m^{-1}s^{-1}}$ [1], but is close to the characteristic times of the back-flow effect [6], i.e. coupling of the director reorientation and mass flow [1]. The other possible reason, ionic currents, does not seem plausible, because the 0.5 ms universal decay behavior does not change when we replace the dc pulse with two subsequent pulses of opposite polarities and half duration.

We found experimentally that the universal decay can be suppressed by applying a long low-amplitude ac pulse that changes the director orientation prior to the high-voltage dc pulse, Fig. 3(b), giving further evidence of the involvement of back-flow and bringing the system closer to the "no backflow" condition for which the theory has been developed. With the suppressed decay, the difference between the response to pulses (i), (e1) and (e2) is very pronounced, Fig. 3(b): (e1) produces a much faster reorientation as compared to (i), while (e2) results in a non-monotonous behavior; (r-e1) and (r-e2) in Fig. 3(b) are well fitted with Eq. (4) as explained above, using the experimental $a$, $\tau = 20$ $\mu$s and $\Gamma$ and the single fitting parameter $\rho = 0.496$.

We further verified the theory in control experiments with a planar cell filled with 5CB, in which $\varepsilon_\parallel$ experiences relaxation with $\tau \approx 50$ ns [7]. Such a short $\tau$ should not lead to any observable DME in our experiments with the typical 1 $\mu$s time of voltage changes. Indeed, as shown in Fig. 4, two (e)-pulses with the same amplitude $|a|=1$ and duration $\tau/\Gamma = 10$ $\mu$s, but of *opposite* polarity, produce *the same* positive torque. This feature is observed regardless of whether the cell is pre-addressed with the ac pulses to suppress the universal decay, Fig. 4(b), or not, Fig. 4(a). Such a behavior is consistent with the non-dispersive character of 5CB in the kHz range and with the classic instantaneous model with dielectric torque quadratic in $\mathbf{E}$; it is clearly different from the behavior of a dispersive NLC that is sensitive to the polarity of the driving pulses, Fig. 2.

**Conclusion.** Theory and experiments above demonstrate that dielectric response in a NLC with dielectric dispersion might be sensitive to the polarity of the applied voltage. The effect is caused by a special "memory" term in the dielectric torque $\mathbf{M}_d(t)$ that is linear in the present field $\mathbf{E}(t)$, in contrast to a regular contribution that is quadratic in $\mathbf{E}(t)$. This feature opens new possibilities for optimization of electrooptical effects in NLCs. For example, we demonstrated that the "switch-off" phase of director reorientation can be accelerated by exponentially decaying short pulses of a proper polarity with the duration determined by $\tau$.

Linear in $\mathbf{E}(t)$ character of the "memory" dielectric torque in a dispersive NLC offers a possibility of interplay with other field effects, such as flexoelectricity, order electricity, surface polarization, etc. Figure 3 suggests that the hydrodynamic processes in the NLC cells are also coupled to the DME, as the relaxation behavior changes when the "universal decay" trend is suppressed. In other words, the observed polarization-sensitive dielectric response of the NLCs opens perspectives for both applied and basic research.

We thank P. Bos for discussions. The work was supported by DOE Grant No. DE-FG02-06ER 46331.

Figure 1. Dielectric dispersion of two NLCs, with $\Delta\varepsilon < 0$ (a) and $\Delta\varepsilon > 0$ (b). The data for $\varepsilon_\parallel$ are fitted by Eq. (1) with $\varepsilon_{l\parallel} = 6.60$, $\varepsilon_{h\parallel} = 3.91$, $\varepsilon_\perp = 7.31$ and $\tau = 33\,\mu s$ (a); $\varepsilon_{l\parallel} = 15.21$, $\varepsilon_{h\parallel} = 8.67$, $\varepsilon_\perp = 7.88$ and $\tau = 20\,\mu s$ (b).

Figure 2. Electrooptic response $I(t)$ of a $\Delta\varepsilon < 0$ NLC in a homeotropic cell driven by dc pulses with an instantaneous (i) and exponentially decaying back edges (e1-e3); "r" stands for "response". The dashed lines show $I(t)$ simulated using Eqs. (4) and (5). The inset shows $I(t)$ over a large time scale.

Figure 3. Electrooptic response $I(t)$ of a $\Delta\varepsilon > 0$ NLC in a planar cell driven by dc pulses with an instantaneous (i) and exponentially decaying back edges (e1, e2). (a) The cell is driven by a dc pulse of duration $310\,\mu s$; (b) the cell is driven by a 5 V, 1 kHz ac pulse of duration 180 ms, followed by a dc pulse of duration $120\,\mu s$; the inset shows a larger time scale. The dashed lines show $I(t)$ simulated using Eqs. (4) and (5).

Figure 4. Electrooptic response $I(t)$ of 5CB in a planar cell driven by dc pulses with an instantaneous (i) and exponentially decaying back edges (e1, e2) of opposite polarity. (a) The cell is driven by a dc pulse of duration $121\,\mu s$; (b) the cell is driven by a 5 V, 1 kHz ac pulse of duration 180 ms (not shown), followed by a dc pulse of duration $80\,\mu s$.

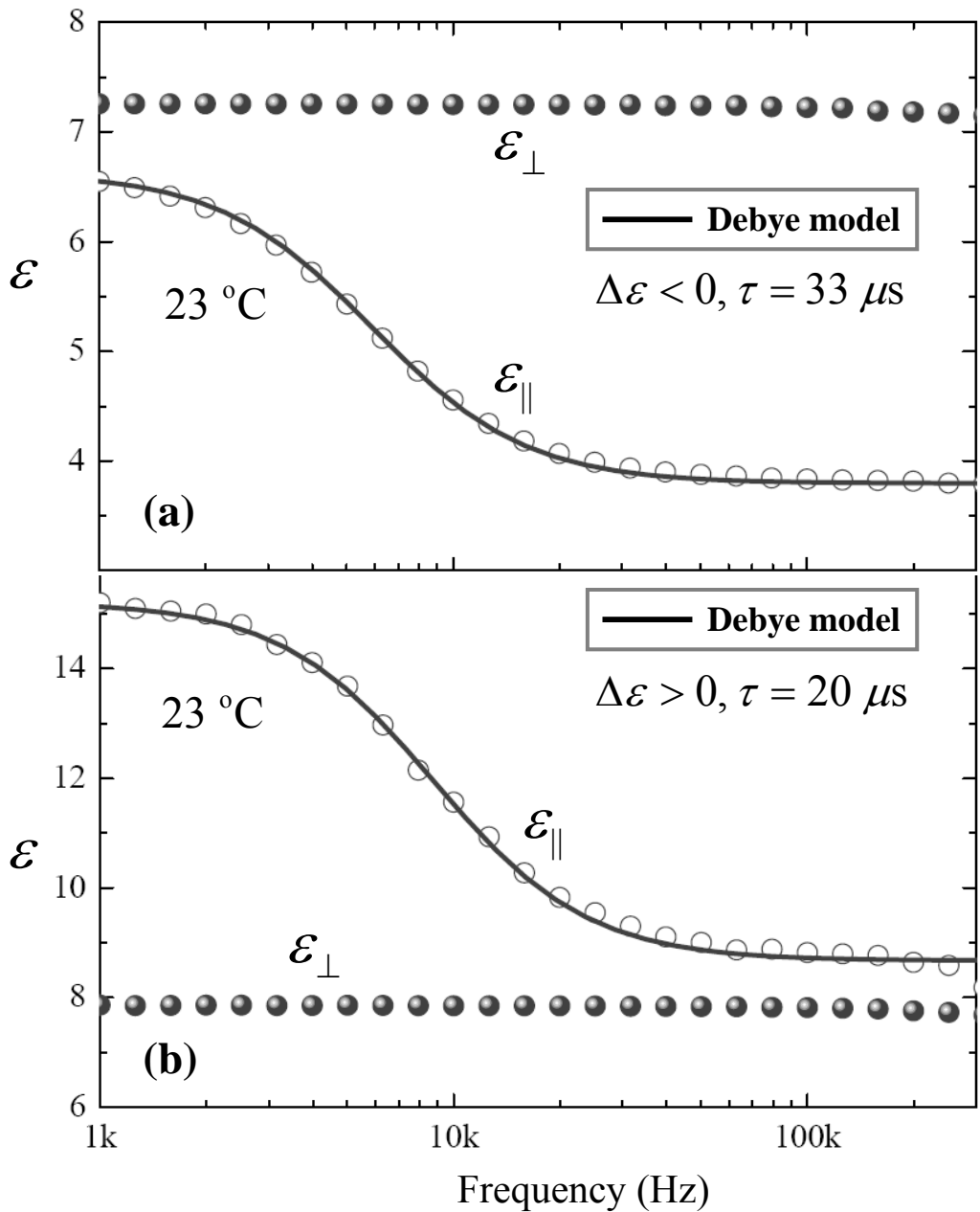

Fig. 1

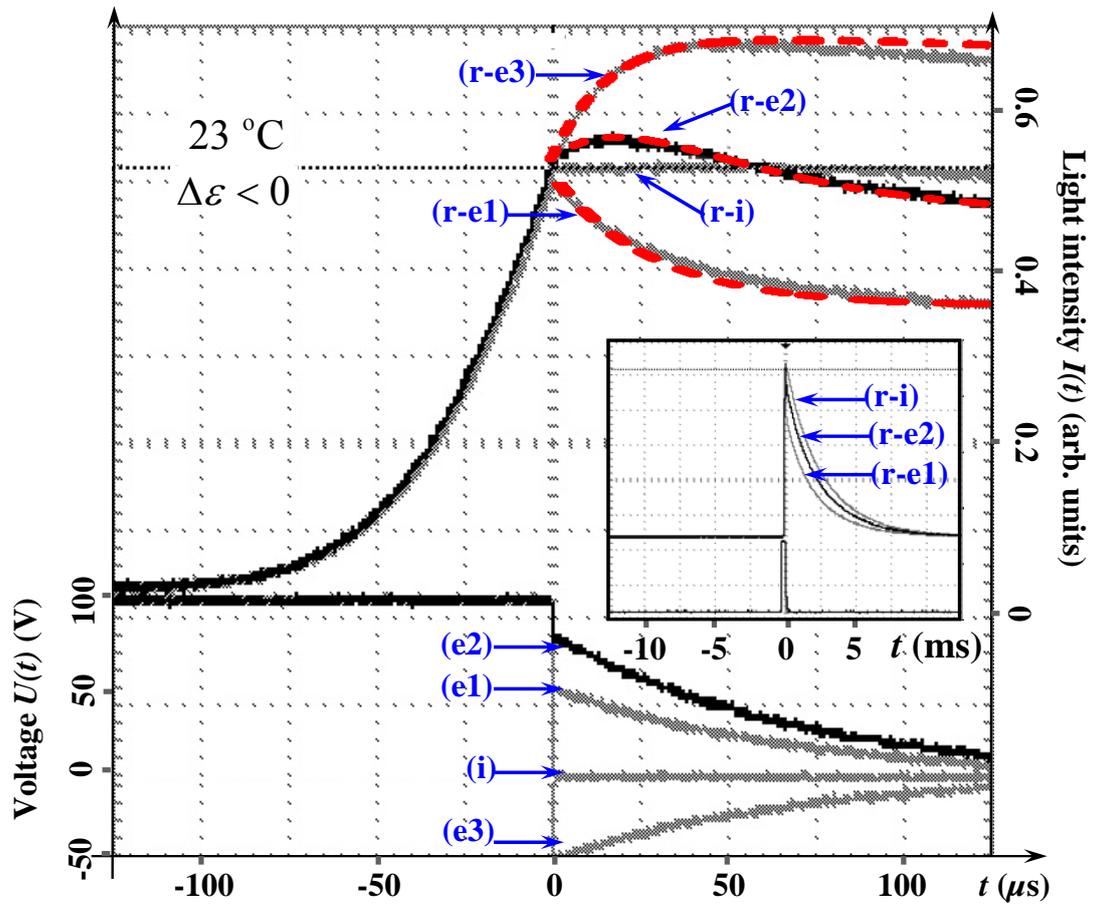

Fig. 2

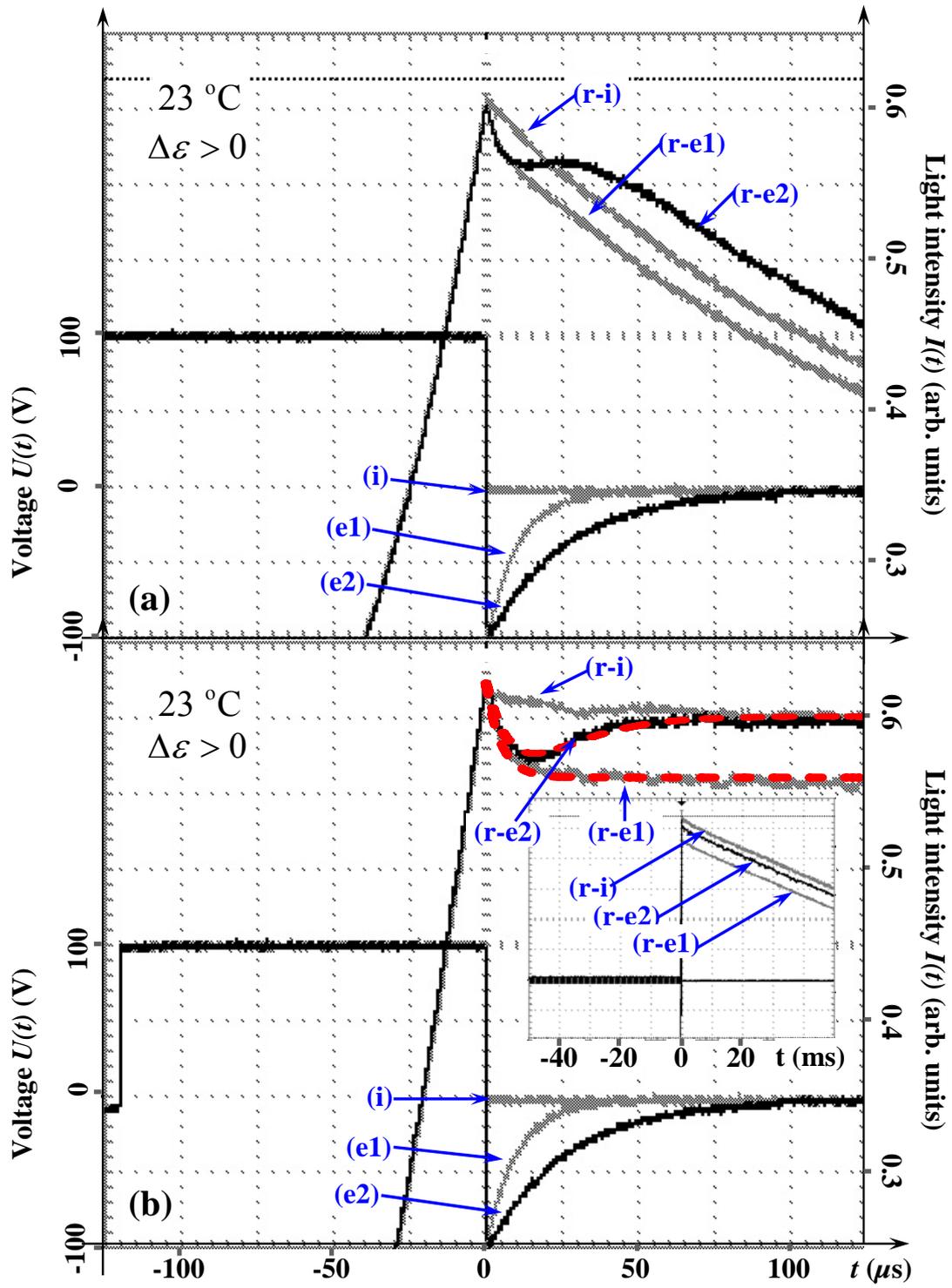

Fig. 3

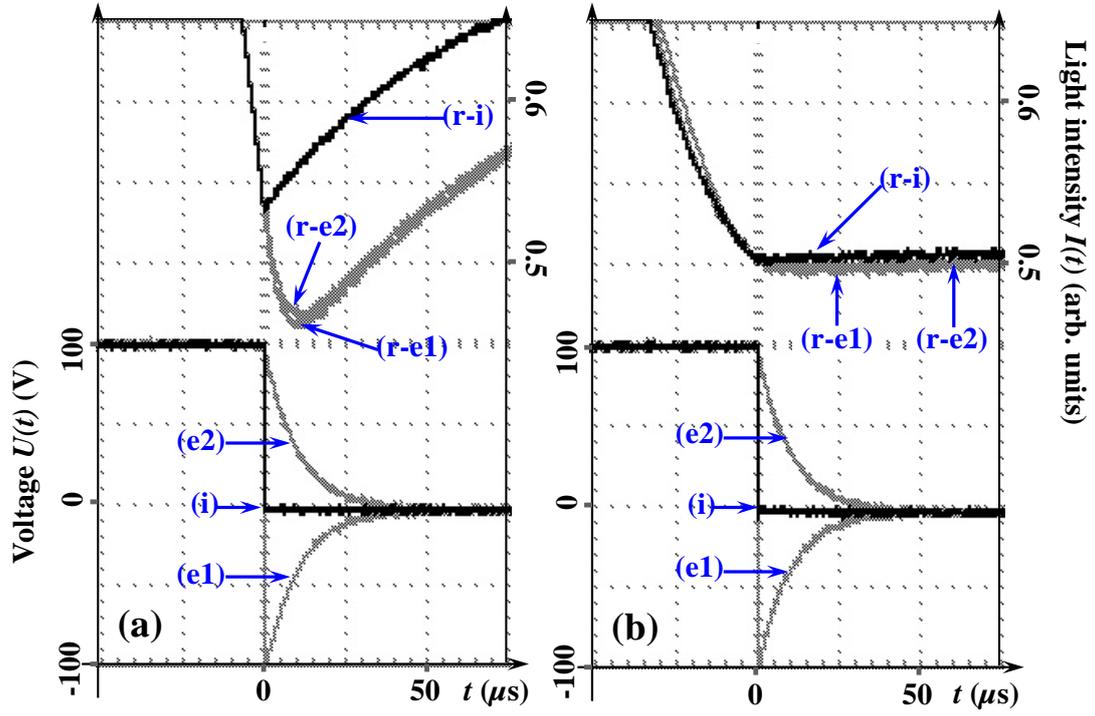

Fig. 4